\documentclass[journal, 12pt, onecolumn]{IEEEtran}
\ifCLASSINFOpdf
\else
\fi

\usepackage{amsfonts, amsmath,amssymb,algorithm,algorithmic,
multirow,xcolor, indentfirst,bm,graphicx, epstopdf, cite,pifont}
\usepackage{multicol}

\definecolor{JHcolor}{RGB}{201,208,201}

\begin{document}
\title{MILLIMETER-WAVE PROPAGATION MODELING AND MEASUREMENTS
FOR 5G MOBILE NETWORKS}
\author{Zhijian Lin,~\IEEEmembership{Member,~IEEE,} Xiaojiang Du,~\IEEEmembership{Senior Member,~IEEE,} \\Hsiao-Hwa Chen*,~\IEEEmembership{Fellow,~IEEE,} Bo Ai*,~\IEEEmembership{Senior Member,~IEEE,} \\Zhifeng Chen,~\IEEEmembership{Senior Member,~IEEE,} and Dapeng Wu,~\IEEEmembership{Fellow,~IEEE}
\thanks{Dr. Zhijian Lin and Prof. Zhifeng Chen are with the Department of
Electrical \& Information Engineering, Fuzhou University, Fuzhou, Fujian, 361000, China,
(e-mail: zlin@fzu.edu.cn; zhifeng@ieee.org).}

\thanks{Prof. Xiaojiang Du is with the Department of Computer and Information Sciences,
Temple University, Philadelphia, PA, 19122, USA, (e-mail: dxj@ieee.org).}

\thanks{Prof. Bo Ai is with the State Key Laboratory of
Rail Traffic Control and Safety, Beijing Jiaotong University, Beijing, 100044, China,  (e-mail: boai@bjtu.edu.cn).}

\thanks{Prof. Hsiao-Hwa Chen is with the Department of Engineering Science,
National Cheng Kung University, Tainan, 70101, Taiwan, (e-mail: hshwchen@mail.ncku.edu.tw).}

\thanks{Prof. Dapeng Wu is with the Department of Electrical \& Computer Engineering, Florida University, Florida, 32611-6130, USA,  (e-mail: dpwu@ieee.org).}

}
\maketitle
\section{ABSTRACT}
Millimeter wave (mmWave) communication is one of the most promising technologies in fifth generation (5G) mobile networks due to its access to a large amount of available spectrum resources. Despite the theoretical potential of a high data rate, there are still several key technical challenges with using mmWave in mobile networks, such as severe pathloss, high penetration loss, narrow beamwidth, etc. Hence, accurate and reliable knowledge of mmWave channel propagation characteristics is essential for developing 5G wireless communication systems. In this article, the fundamental characteristics of mmWave are first presented. Then, two main channel modeling methods are discussed. Finally, in order to investigate the channel characteristics at the mmWave band, measurement campaigns using three different large-scale array topologies are carried out and the typical channel parameters are extracted and analyzed.
%
\section{Introduction}

According to the estimate presented in \cite{1}, it is expected that as many as 50 billion devices will be connected to each other around 2020. Mobile data traffic will increase up to 1000-fold over the next decade, as predicted in \cite{2}. With the explosive increase in mobile service and user demands, the increasing number of connected devices will put significant pressure on the existing wireless communication systems. In order to get these challenges under control, the wireless industry is moving toward its fifth generation (5G) of cellular technology, which is expected to be deployed by 2020, to increase capacity and improve energy efficiency, cost, as well as spectrum utilization. Since 2013, many national-level 5G research organizations and projects have been created to achieve the technical targets. In 2015, the International Telecommunications Union Radio Communications Sector (ITU-R) officially named 5G as IMT-2020, and released recommendation on its framework and overall objectives.

With the maturing of 4G standardization and the ongoing worldwide deployment of 4G cellular networks, technological research on 5G networks has emerged in both academia and industrial communities. What will the 5G networks look like? It is widely agreed that 5G networks can provide the following key performance indicators: 10 Gbit/s peak data rate, 3x spectrum efficiency, 100x network energy efficiency, and 1 ms over-the-air latency \cite{3}. In order to achieve the magnificent objectives aforementioned, some promising wireless technologies that can enable 5G wireless networks to fulfill performance requirements should be developed. As shown in Figure 1, In-band full-duplex (IBFD), device-to-device (D2D) communication, massive multiple-input multiple-output (MIMO), and millimeter wave (mmWave) are the key ingredients contributing to the capacity increase of 5G \cite{4}.

\begin{figure}[!t]
\centering
\includegraphics[width=0.6\columnwidth]{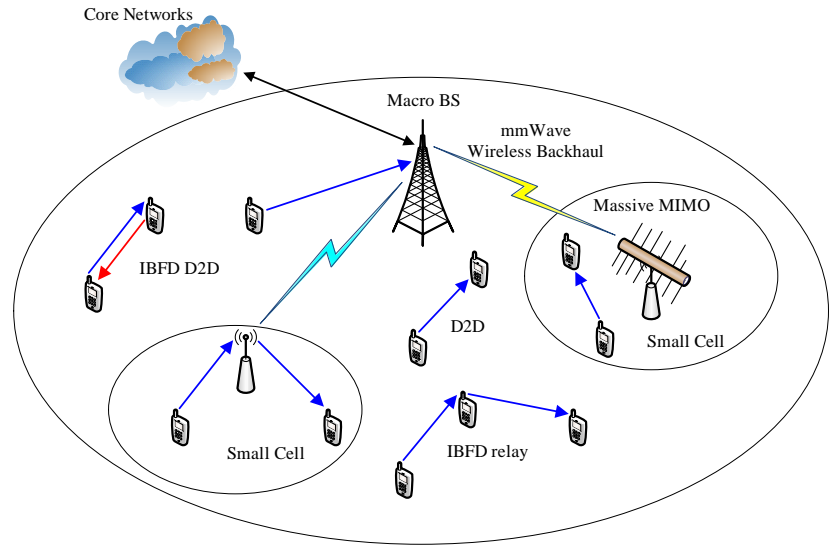}
\caption{Illustration of promising technologies in 5G networks.
} \label{Figure :1}
\end{figure}

It can be observed that the current cellular network mainly operates in frequency bands below 3 GHz, and the total licensed spectrum used today is approximately 1 GHz. As most of the frequency bands below 3GHz are occupied and the air-interface spectrum efficiency has been approaching its capacity limit \cite{5}, the attention on acquiring new spectrum for 5G networks has shifted to frequency bands above 3 GHz. Among the promising technologies of 5G system, we know that a much higher data rate and a much larger system capacity can be achieved by using mmWave. Hence, mmWave communication, which can take advantage of a large amount of available bandwidth, is widely considered in 5G mobile networks. Although the wireless and networking issues have been widely studied \cite{6,7,8,9,10,11}, most of the existing channel characteristics and models are not feasible in the mmWave band due to the fact that the channel propagation characteristics are different from those of frequency bands below 3 GHz. Hence, fundamental knowledge of mmWave channel propagation characteristics, including accurate and reliable channel models, is vital for developing 5G wireless communication systems. Over the past few years, large efforts have been devoted to mmWave communications, where measurements and models of channels under various scenarios have been carried out by many companies and research groups \cite{12,13,14,15}. In this article, we mainly focus on presenting the channel modeling methods and measurement campaigns for mmWave band.

The remainder of this article is organized as follows. The fundamental characteristics of mmWave are discussed first, and the descriptions of channel modeling methods for mmWave band are provided. Furthermore, a channel measurement campaign using three different massive MIMO array topologies are presented. Finally, the article is concluded with a summary and a brief discussion of future work.

\section{FUNDAMENTAL CHARACTERISTICS OF MMWAVE}

MmWave communication, with wavelength on the order of millimeter, has several fundamental characteristics, including a large amount of bandwidth, short wavelength, severe pathloss, high penetration loss, narrow beamwidth, etc. In the following, we will give brief discussion about them.

\subsection{LARGE AMOUNT OF BANDWIDTH}

Currently, the total available bandwidth for mobile networks including 2G, 3G, and 4G is smaller than 780 MHz, which is not sufficient for high data rate requirements from various devices. Compared to the traditional communication with microwave, one of the major benefits of mmWave communications is the significantly larger bandwidth. Even though there are some unfavorable bands, such as 57-64 GHz and 164-200 GHz, which are easily absorbed by oxygen and water vapor, respectively, the suitable bandwidth for mmWave communications can still be more than 150 GHz, and more than 150 Gbps can be achieved on the whole spectrum.

\subsection{SHORT WAVELENGTH}

Compared to microwave signal, mmWave signal has much shorter wavelength, which is on the order of millimeter. Hence, MIMO/massive MIMO is considered essential for mmWave communications, since the short wavelength at mmWave frequencies is beneficial to fit large numbers of half-wavelength spaced antennas into a small area. A large-scale antenna system has the capability of greatly improving spectral efficiency. The combination of mmWave and massive MIMO has the potential to dramatically improve wireless access and throughput performance.

\subsection{PROPAGATION LOSS}

Generally, propagation loss can be interpreted as both pathloss and penetration loss. Under the assumption of line-of-sight (LOS), the free space pathloss is proportional to the square of the carrier frequency according to the Friis transmission formula. Since the frequency range of mmWave is from 26.5 GHz to 300 GHz, the propagation loss is much higher than that in the microwave band, for example, it is 28 dB higher at 60 GHz than at 2.4 GHz \cite{16}. However, a high-gain directional antenna can be applied to compensate for the large propagation loss. Although tremendous pathloss is one of the disadvantages of mmWave, the positive side is that the combination of mmWave with device-to-device (D2D) communication has relatively low multi-user interference due to the high propagation loss and the utilization of directional antennas. This combination can support huge amounts of concurrent D2D links, such that the network capacity can be further improved. In addition, a higher security against eavesdropping and jamming can be achieved.

In non-line-of-sight (NLOS) scenarios, the penetration loss is typically larger at higher frequencies. For example, the attenuation is about 178 dB for a brick wall and 20 dB for a painted board at 40 GHz. Hence, it is difficult to cover indoor spaces with mmWave nodes that are deployed outdoors. For indoor users connecting with outdoor base stations (BSs), the signals have to go through the building walls. This propagation could suffer a very high penetration loss, which significantly degrades the data rate, spectral efficiency, and energy efficiency. The higher the carrier frequency is, the worse the situation is. Therefore, outdoor and indoor scenarios are expected to be separated in the future 5G cellular architecture.

\section{CHANNEL MODELING IN MMWAVE BAND}

To the best of the authors＊ knowledge, mmWave signal has stronger reflectivity when interacting with metal, glass, etc., and is more easily absorbed by air, rain, etc., compared with the signal at lower frequency bands. Moreover, its diffraction ability is also lower than that in today＊s wireless networks. It is essential to obtain the fundamental knowledge of the mmWave channel propagation characteristics for developing 5G wireless communication systems. Emerging mmWave technology will require new channel models to facilitate the real system design in 5G.
Generally, most channel modeling methods can be classified into two main categories: analytical modeling and ray-tracing based modeling. The analytical modeling can be expressed by a fixed set of parameters and ray-tracing based modeling relies on finding the signal paths in the environment. The ray-tracing based method will be more suitable for mmWave since high frequencies cause more reflection and less diffraction.

\subsection{ANALYTICAL MODELING METHOD}

Based on the data of measurements or statistical characteristics of the scenario, the corresponding statistical parameters such as number of paths, root-mean-square (RMS) delay spread, path loss, and shadowing of the propagation channel can be obtained and we refer to it as the analytical modeling method. This method can be expressed by a fixed set of parameters without considering the details of the environment. Hence, this may cause inaccuracy of the analysis result in an anisotropic radio environment.

\subsection{RAY-TRACING METHOD}

In mmWave wireless communication systems, directional antenna and beamforming techniques are generally utilized to find the optimal path with minimum loss to reach the users and keep the interference as low as possible. However, it is difficult to rapidly adapt to the main power paths when the environment is changing. One promising solution is the ray-tracing method. Based on the prior knowledge of the geometry of the surrounding and the transceiver in the given scenario, the ray-tracing method can be used to identify the potential powerful transmission paths, and this will be helpful in dealing with the changing environment. In addition, in the scenario of massive MIMO, a ray-tracing based method can also be used to obtain the channel characteristics of more complex scenarios with a larger number of array elements of antennas, and this will be useful for the design of a real mmWave system in indoor or outdoor environments in the foreseeable future.

\section{SCENARIO OF MEASUREMENTS IN MMWAVE CHANNEL}

In order to investigate the channel characteristics at mmWave frequency bands, an indoor massive MIMO channel measurement campaign was conducted at 26 GHz. Three different large-scale array topologies were considered. Based on the measurement data, the typical channel parameters are extracted and analyzed.

\subsection{MEASUREMENT SYSTEM}

The channel sounder is developed with the support of a high-performance vector signal generator and a broadband vector signal analyzer. Three different virtual array topologies, i.e. a 64-element linear array, a 64-element planar array (8$\times$8), and a 128-element planar array (8$\times$16) were constituted with the help of a three-dimensional (3D) mechanical positioner. The carrier frequency is set to be 26 GHz, and the measurement bandwidth is 200 MHz. Both the Tx and Rx antennas are omnidirectional vertical polarized antennas.

\subsection{MEASUREMENT ENVIRONMENT}

The channel measurement campaign for massive MIMO was conducted in a lecture hall in NO.9 Teaching Building, Beijing Jiaotong University, China, as shown in Figure 2. The lecture hall can accommodate up to 300 people, which is considered to be a typical application scenario for 5G. The transmitter (Tx) is placed on the stage of the hall, and the receiver (Rx) is located in the aisle between the seats. The distance between the Tx and the Rx is around 5.0 m. The Tx antenna is mounted on the adjustable shelf of the positioner, and the Rx antenna is fixed on the a pole. The height of the Tx and Rx antenna is 2.5 m and 2.0 m, respectively. The hall has a dimension of 20.1 m $\times$ 20.2 m $\times$ 4.5 m (length $\times$ width $\times$ height). The grounds, walls, and ceiling are made of concrete, the seats are covered with cloth, and the tables in front of the seats are made of wood. The surface of the stage is made with wood and covered with cloth. On the walls at both sides of the hall, there are some widows and albums, which are mainly made of glass. During the measurements, there were no people or any other obstacles moving in the environment. Thus, the channels are considered to be quasi-static.

\subsection{DATA PROCESSING}

The sounder system recorded the in-phase and quadrature (I/Q) data during the measurements and the channel impulse response (CIR) for each subchannel is calculated. The average power delay profile (APDP) is obtained from the CIR. When calculating the typical channel parameters, such as RMS delay spread, channel gain, and K factor, the data from the three arrays are combined as a whole. In addition, the angular characteristics of the channel are investigated. The multipath components (MPCs) in the environment are extracted by using the space-alternating generalized expectation maximization algorithm (SAGE) \cite{17}.

\section{MEASUREMENT RESULTS AND ANALYSIS}

\subsection{APDP AND RMS DELAY SPREAD}

The channel dispersion in the delay domain can be characterized by the RMS delay spread, which is calculated based on the obtained APDP as follows \cite{18}, where $\tau_{p}$ denotes the $p$-th delay bin of the corresponding APDP. Same with \cite{19}, where a threshold of 6 dB beyond the noise floor is set to remove the noise component in the APDP. Figure 3 illustrates the APDPs along the 64-element linear array and the cumulative distribution functions (CDFs) of the RMS delay spread in the measurements. It is found that the RMS delay spread can fit well with a Lognormal distribution. At 26 GHz, the RMS delay spread is generally less than 18 ns. This result is smaller than those at frequency below 6 GHz in \cite{20}.

\subsection{CHANNEL GAIN ANG K FACTOR}

The channel gain and the K factor reflect the strength of the channel response and its ratio of the LOS component and the NLOS component, respectively. The channel gain G and the K factor $K$ can be both obtained from the CIR. Figure 4 shows the CDFs of the channel gain and K factor. It can be seen that both of these two parameters can fit well with Normal distributions and the K factor is mostly larger than 0 dB, since the LOS condition is held in our measurements.

\begin{figure}[!t]
\centering
\includegraphics[width=0.6\columnwidth]{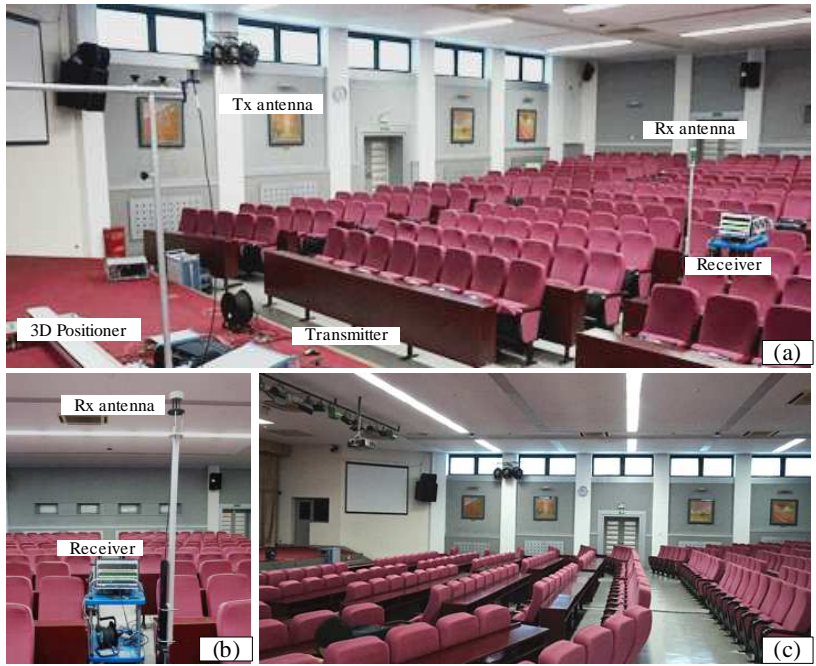}
\caption{Photos of measurement environment. (a) Locations of Tx and Rx.
(b) Receiver and Rx antenna. (c) Walls of the lecture hall.
} \label{Figure :2}
\end{figure}

\subsection{SPATIAL NONSTATIONARITY}

One of the important features of massive MIMO channels is the spatial non-stationarity, which has already been observed in our preliminary work in the same environment \cite{21}. We investigate this characteristic in the angular domain at 26 GHz in this article, by using the SAGE algorithm. When dealing with the data of 64-element linear array, a sliding window of 10 neighboring elements is used. Figure 5(a) shows the estimated azimuth of departure (AOD) of MPCs along the 64-element linear array, and Figure 5(b) shows the estimated AOD and elevation of departure (EOD) for a 128-element planar array. From Figure 5(a), it can be observed that some of the MPCs only are observable by parts of the array. We refer to this phenomenon as the birth-death (BD) process of the MPCs in the array dimension. The LOS component (i.e. the strongest MPC) can be clearly observed, and the power of the dominant MPCs varies along the large-scale array. It should be noticed that the spatial non-stationarity of the channel can be also observed in the delay domain, as shown in Figure 3(a).

\begin{figure}[!t]
\centering
\includegraphics[width=0.6\columnwidth]{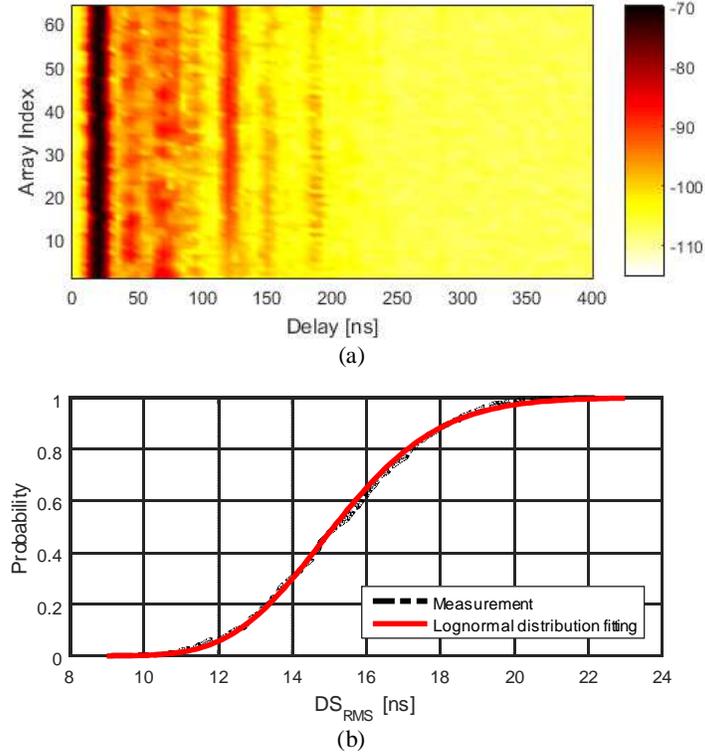}
\caption{APDPs and calculated RMS delay spread. (a) APDPs along the 64-element linear array. (b) CDFs of RMS delay spread, together with the Lognormal distribution fitting.
} \label{Figure :3}
\end{figure}

\subsection{SCATTERER IDENTIFICATION}

We identify the scatterers by directly relating the SAGE estimates to the physical objects in the measurement environment. It can be observed from Figure 5 that the NLOS components are generally much weaker than the LOS component at 26 GHz. The LOS component, as marked by {\large\ding{192}} in Figure 5(a), carries most of the transmitted power. The MPCs indicated by {\large\ding{193}} and {\large\ding{195}} are probably the reflections from the seats near the both sides of the aisle. The MPCs indicated by {\large\ding{194}} and {\large\ding{196}} are probably the reflections from the walls at both sides of the hall. For SAGE estimates for the 128-element planar array, the cluster around $-60^{\circ}$ and $40^{\circ}$ of AOD are created by the walls at both sides of the hall. In a wide angle range around the LOS component, there are numbers of MPCs components, which are probably the reflections from the seats near the both sides of the aisle. In addition, a few strong MPCs (i.e. around $50^{\circ}$ and $70^{\circ}$ of AOD) are observed; they are probably from the wooden tables in front of the seats.

\begin{figure}[!t]
\centering
\includegraphics[width=0.6\columnwidth]{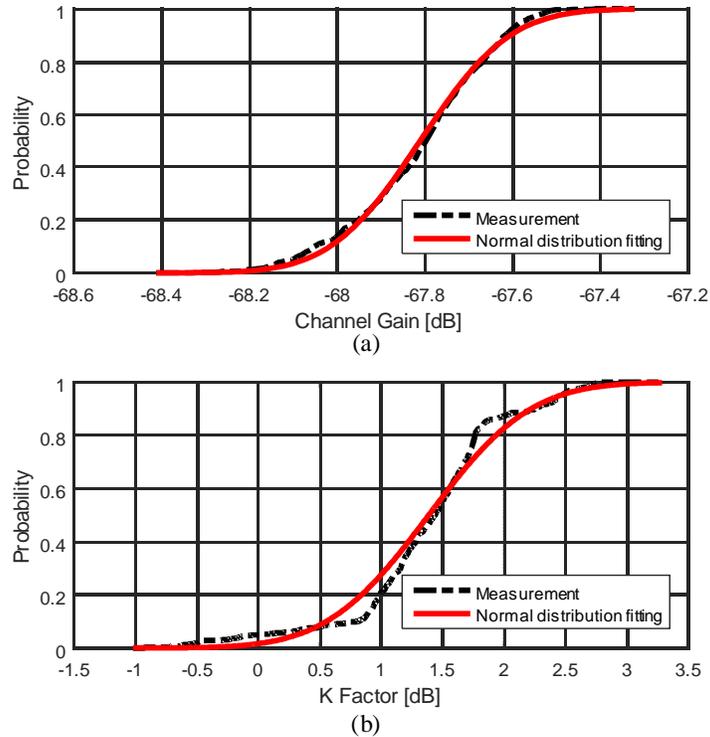}
\caption{Calculated channel gain and global K factor. (a) CDFs of channel gain, together with the Normal distribution fitting. (b) CDFs of global K factor, together with the Normal distribution fitting.
} \label{Figure :4}
\end{figure}

\begin{figure}[!t]
\centering
\includegraphics[width=0.6\columnwidth]{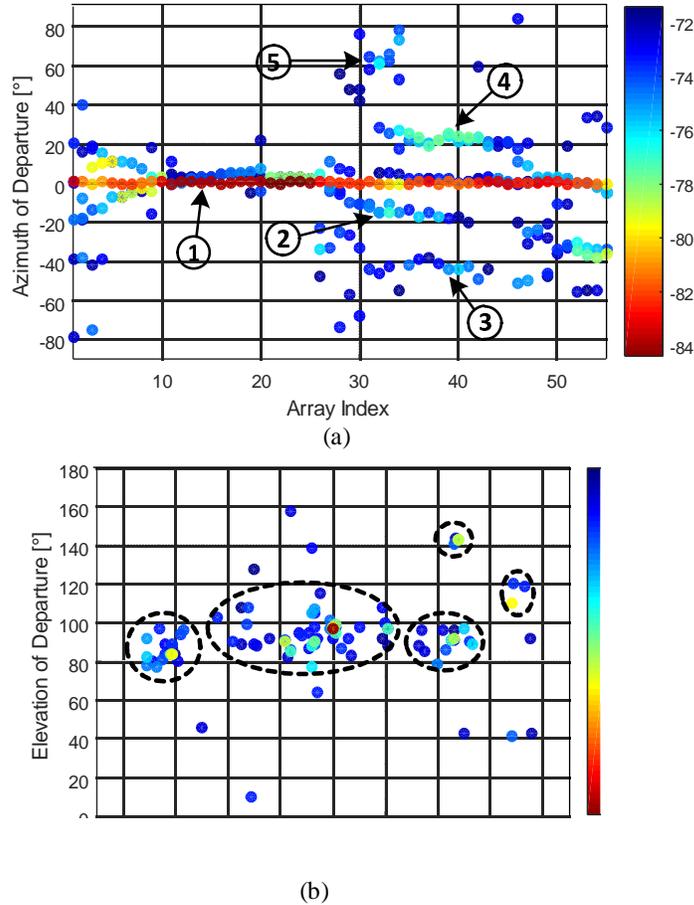}
\caption{SAGE estimates. Each solid dot represents an MPC, and the dB-scaled power of the MPCs is represented by color. (a) Estimated AOD of MPCs along the 64-element linear array. (b) Estimated AOD and EOD of the MPCs for the 128-element planar array.
} \label{Figure :4}
\end{figure}

\section{CONCLUSION}

In this article, the fundamental characteristics of mmWave and two main channel modeling methods are discussed. In addition, this article presents a comprehensive research on the indoor mmWave MIMO channel. To start, the measurement system and environment for three different large-scale array topologies are provided. After the measurement and data processing, analysis about the measurement results is presented including APDP, RMS delay spread, channel gain, K factor, spatial non-stationarity, and scatterer identification.

To grasp the propagation and channel characteristics, measurement campaigns at different scenarios need to be conducted. Our future work will continue to bring forward the measurements of massive MIMO channel at a wide frequency band under different scenarios, such as subway station, theatre, urban small cells, etc.


\begin{thebibliography}{15}

\bibitem{1} Ericsson, ※More Than 50 Billion Connected Devices,§ White Paper [online] http://www.ericsson.comires/docs/whitepapers/wp-50billions.pdf, 2011.
\bibitem{2} Cisco, ※Cisco Visual Network Index: Global Mobile Traffic Forecast Update,§ 2013.
\bibitem{3} J. G. Andrews et al., ※What will 5G be?§ IEEE J. Sel. Areas Commun., vol. 32, no. 6, pp. 1065每1082, Jun. 2014.
\bibitem{4} Q. C. Li, H. Niu, A. T. Papathanassiou, and G. Wu, ※5G network capacity: key elements and technologies,§ IEEE Veh. Technol. Mag., vol. 9, no. 1, pp. 71-78, 2014.
\bibitem{5} T. S. Rappaport, S. Sun, R. Mayzus, H. Zhao, Y. Azar, K. Wang and F. Gutierrez, ※Millimeter wave mobile communications for 5G cellular: It will work!§ IEEE access, vol. 1, pp. 335-349, 2013.

\bibitem{6} H. Zhang, S. Chen, X. Li, H. Ji, and X. Du, ※Interference Management for Heterogeneous Network with Spectral Efficiency Improvement,§ IEEE Wireless Communications Magazine, Issue 2, Vol. 22, pp. 101-107, April 2015.
\bibitem{7} H. Zhang, Q. Zhang, and X. Du, ※Toward Vehicle-Assisted Cloud Computing for Smartphones,§ IEEE Transactions on Vehicular Technology, Issue 12, Vol.64, pp. 5610-5618, Dec. 2015.
\bibitem{8} X. Du, M. Zhang, K. Nygard, S. Guizani, and H. H. Chen, ※Self-Healing Sensor Networks with Distributed Decision Making,§ International Journal of Sensor Networks, Vol. 2, Nos. 5/6, pp. 289 每298,  2007.
\bibitem{9} L. Wu, and X. Du, ※MobiFish: A Lightweight Anti-Phishing Scheme for Mobile Phones,§ in Proc. of the 23rd International Conference on Computer Communications and Networks (ICCCN), Shanghai, China, August 2014.
\bibitem{10} X. Du, Y. Xiao, M. Guizani, and H. H. Chen, ※An Effective Key Management Scheme for Heterogeneous Sensor Networks,§ Ad Hoc Networks, Elsevier, Vol. 5, Issue 1, pp 24每34, Jan. 2007.
\bibitem{11} Y. Xiao, X. Du, J. Zhang, and S. Guizani, ※Internet Protocol Television (IPTV): the Killer Application for the Next Generation Internet,§ IEEE Communications Magazine, Vol. 45, No. 11, pp. 126每134, Nov. 2007.

\bibitem{12} Q. Wang, D. W. Matolak, and B. Ai, ※Shadowing Characterization for 5 GHz Vehicle-to-Vehicle Channels,§ IEEE Trans. on Veh. Tech., 2017.
\bibitem{13} S. Rangan, T. S. Rappaport, and E. Erkip, ※Millimeter-wave cellular wireless networks: Potentials and challenges,§ Proc. IEEE, vol. 102, no. 3, pp. 366每385, Mar. 2014.
\bibitem{14} L. Huang, G. Zhu, and X. Du, ※Cognitive Femtocell Networks: An Opportunistic Spectrum Access for Future Indoor Wireless Coverage,§ IEEE Wireless Communications Magazine, Vol. 20, Issue 2, pp. 44 每 51, April, 2013.
\bibitem{15} O. H. Koymen, A. Partyka, S. Subramanian, and J. Li, ※Indoor mmwave channel measurements: Comparative study of 2.9 GHz and 29 GHz,§ in Proc. IEEE Global Commun. Conf. (GLOBECOM), pp. 1每6, Dec. 2015.
\bibitem{16} A. Maltsev, R. Maslennikov, A. S evastyanov, A. Khoryaev, and A. Lomayev, ※Experimental investigations of 60 GHz WLAN systems in office environment,§ IEEE J. Sel. Areas Commun., vol. 27, no. 8, 2009.
\bibitem{17} C. C. Chong, D. I. Laurenson, C. M. Tan, S. McLaughlin, M. A. Beach, and A. R. Nix, ※Joint detection-estimation of directional channel parameters using the 2-D frequency domain SAGE algorithm with serial interference cancellation,§ IEEE International Conference on ICC, vol. 2, pp. 906每910, 2002.
\bibitem{18} A. F. Molisch, Wireless Communications, 2nd ed. Wiley Publishing, 2011.
\bibitem{19} A. F. Molisch and M. Steinbauer, ※Condensed parameters for characterizing wideband mobile radio channels,§ International Journal of Wireless Information Networks, vol. 6, no. 3, pp. 133每154, 1999.
\bibitem{20} J. Li, B. Ai, R. He, K. Guan, Q. Wang, D. Fei, Z. Zhong, Z. Zhao, D. Miao, and H. Guan, ※Measurement-based characterizations of indoor massive MIMO channels at 2 GHz, 4 GHz, and 6 GHz frequency bands,§ in 2016 IEEE 83rd Vehicular Technology Conference (VTC Spring), May 2016, pp. 1每5.
\bibitem{21} B. Ai, K. Guan, R. He, J. Li, G. Li, D. He, Z. Zhong, and K. M. S. Huq, ※On indoor millimeter wave massive MIMO channels: Measurement and simulation,§ IEEE Journal on Selected Areas in Communications, vol. 35, no. 7, pp. 1678每1690, July 2017.


\end{thebibliography}
\end{document}